\documentclass{aa}  
\usepackage{graphicx}
\usepackage{txfonts}
\begin{document}
  \title{Modelling element distributions in the atmospheres 
         of magnetic Ap stars}

  \author{G.~Alecian\inst{1} \and M.J. Stift\inst{2}\inst{,1} }

  \offprints{G. Alecian}
 
  \institute{LUTH, Observatoire de Paris, CNRS, Universit{\'e} Paris Diderot, 
             5 Place Jules Janssen, 92190 Meudon, France\\
             \email{georges.alecian@obspm.fr}
             \and
             Institut f{\"u}r Astronomie (IfA), Universit{\"a}t Wien,
             T{\"u}rkenschanzstrasse 17, A-1180 Wien, Austria\\
             \email{stift@astro.univie.ac.at}
             }

   \date{Received 1 June 2007; Accepted 31 Jul 2007}

   \abstract
     {In recent papers convincing evidence has been presented for chemical
     stratification in Ap star atmospheres, and surface abundance maps have been
     shown to correlate with the magnetic field direction. Radiatively driven
     diffusion, which is known to be sensitive to the magnetic field strength
     and direction, is among the processes responsible for these
     inhomogeneities.}
     {Here we explore the hypothesis that equilibrium stratifications -- such
     that the diffusive particle flux is close to zero throughout the atmosphere
     -- can in a number of cases explain the observed abundance maps and
     vertical distributions of the various elements.}
     {An iterative scheme adjusts the abundances in such a way as to achieve either
     zero particle flux or zero effective acceleration throughout the
     atmosphere, taking strength and direction of the magnetic field
     into account.}
     {The investigation of equilibrium stratifications in stellar atmospheres
     with temperatures from 8500 to 12000\,K and fields up to 10\,kG reveals
     considerable variations in the vertical distribution of the 5 elements
     studied (Mg, Si, Ca, Ti, Fe), often with zones of large over- or
     under-abundances and with indications of other competing processes (such
     as mass loss). Horizontal magnetic fields can be very efficient in helping
     the accumulation of elements in higher layers.}
     {A comparison between our calculations and the vertical abundance profiles
     and surface maps derived by magnetic Doppler imaging reveals that
     equilibrium stratifications are in a number of cases consistent with the
     main trends inferred from observed spectra. However, it is not clear
     whether such equilibrium solutions will ever be reached during the
     evolution of an Ap star.}


\titlerunning{Element distributions in Ap stars}
\maketitle

\section{Introduction}

Atomic diffusion in stars, when efficient enough to overcome mixing
processes, leads to inhomogeneous abundance distributions of chemical 
elements. This is probably what happens in the atmospheres of upper 
main-sequence, chemically peculiar (CP) stars which exhibit a wide
variety of strong abundance anomalies, since the outer layers of these 
stars can be considered stable enough to allow element diffusion
to take place. A considerable number of papers, starting with Michaud 
(\cite{mic70}), have examined to what extent the diffusion model is able to 
explain the observed anomalies. Often, but not always, these anomalies 
appear to be quite well correlated with the respective radiative 
accelerations (frequently the leading contribution to the diffusion velocity) 
of each element. However, the CP phenomenon involves so many complex 
processes that a direct comparison of observed apparent 
abundance anomalies and calculated radiative accelerations will not be
sufficient for fully describing the build-up of abundance peculiarities. 
On the one hand, abundance determinations in the past did not take
stratification of the chemical elements into account; on the other, these
stratifications are built up by a time-dependent, non-linear diffusion 
process which is quite sensitive to magnetic fields and to macroscopic 
motions (residual turbulence or stellar wind for instance) and depends 
not only on radiative accelerations.

Despite the fact that full modelling of atmospheres of CP stars is still 
out of reach, notable progress has been made in the field of diffusion.
A first study addressing the special behaviour of silicon in magnetic 
atmospheres was carried out by Vauclair et al. (\cite{vhp79}), followed by
the quantitative modelling of Si stratification by Alecian \& Vauclair 
(\cite{alv81}). A theoretical prediction of manganese accumulation in hot CP
stars was proposed by Alecian \& Michaud (\cite{alm81}), and a first attempt 
at detecting such a stratification was carried out by Alecian (\cite{ale82}) 
in the HgMn star $\upsilon$\,Her using a method based on the curve of growth 
of Mn resonance lines. Abundances of iron peak elements in several HgMn stars 
were analysed in detail by Smith \& Dworetsky (\cite{smi93}). 
In their study, these authors used what were state 
of the art methods at that time, applying schematic corrections for 
the chemical stratifications. Within the framework of this approximate 
treatment, they found that their results were in excellent agreement 
with the predictions of the diffusion model. A detailed study of the
stratifications of several metals in the magnetic star 53 Cam, assuming
the presence of a stellar wind, is due to Babel \& Michaud (\cite{bam91}) 
and to Babel (\cite{bam92}). Their main conclusion for that particular star 
was that diffusion alone cannot account for the observations and that 
more sophisticated models have to be developed, including mass loss 
confined by magnetic fields. These early studies were limited by 
technical constraints, such as insufficient computing power, and by the
lack of atomic and observational data. Therefore no firm conclusions
could be reached concerning the chemical stratifications produced by 
diffusion processes. Fortunately, the situation has improved 
drastically. Thanks to high performance detectors, inhomogeneous element 
distributions appear to be established beyond reasonable doubt; see
e.g. Kochukhov et al. (\cite{koc04}) for horizontal distributions in the 
magnetic atmosphere of 53 Cam and Kochukhov et al. (\cite{koc06}) for
vertical distributions in the Ap star HD\,133792. Significant progress
has also been made in modelling, since self-consistent 
atmospheric models for non-magnetic stars, including abundance 
stratifications compatible with the amount of elements which can be 
supported by radiative forces, are in an advanced stage of development
(Hui-Bon-Hoa et al. \cite{hbh02}).

In the present paper, we attempt to advance one further step on the 
long path towards the complete modelling of the migration process of 
chemical elements, by looking for an equilibrium solution to the
stratification of metals in magnetic atmospheres. We rely on the 
physics and methods presented in previous papers, where we have 
computed the Zeeman amplification of radiative accelerations in 
detail (Alecian \& Stift \cite{als04}), and studied diffusion 
velocities in magnetic atmospheres (Alecian \& Stift, \cite{als06}). 
The present work still assumes LTE and the temperature/pressure 
structure of the atmosphere is computed with solar abundances 
(ATLAS9, Kurucz, \cite{kur93}). The CARAT code is  used in the improved
version discussed by Alecian \& Stift (\cite{als06}). It carries out full 
opacity sampling of Zeeman split spectral lines, determines the 
radiative flux by solving the polarised equation of radiative 
transfer, includes radiative accelerations due to bound-free 
transitions, and takes the redistribution of momentum among ions 
into account. In Section \ref{sec:buildup} we discuss theoretical 
aspects of the element stratification process and in Section 
\ref{sec:numerics} we present the numerical method used to obtain 
solutions for equilibrium stratifications. In Section 
\ref{sec:resdisc} we discuss our results in view of recent observations.

\section{Build-up of abundance stratifications}
\label{sec:buildup}

In a multicomponent gas with anisotropic structure, pressure gradient, 
etc., each component experiences forces specific to its properties and 
diffuses with respect to the others. In such a situation, there is no
reason why the gas should be homogeneous (except if mixing motions
enforce homogeneity). This is what happens in stars in places where 
mixing motions are not strong enough to erase the effects of the 
ineluctable tendency of chemical species to migrate. The change in the 
local chemical composition is described by the continuity equation and
requires knowledge of the average velocity of each component. Several
factors make modelling this process difficult and the results 
problematic. The first difficulty arises from the very low values of 
the diffusion velocities which make their effects on stratification 
very sensitive to any uncertainties in macroscopic motions. The second 
difficulty comes from the complexity of the diffusion process itself, 
which is described by high-order terms of the Boltzmann equation and 
which requires quite a number of approximations before usable expressions 
can be written down. The third difficulty comes from the estimation of 
radiative acceleration, the leading component of the diffusion 
velocity in outer stellar layers. The accurate calculation of radiative 
accelerations is computationally quite expensive; accelerations depend 
in a non-linear way on the concentrations of the elements. Since the 
diffusion process changes the local concentration of elements, 
radiative accelerations are not constant throughout the process. 
Moreover, as discussed in Alecian \& Stift (\cite{als06}), there are still 
theoretical uncertainties concerning the determination of the total 
radiative acceleration (related to the redistribution of momentum 
among ions of a given element). The final difficulty is numerical and 
concerns the solution of the continuity equation. In the following 
subsections, we discuss some of these points in more detail.

\subsection{Some theoretical considerations}
\label{subsec:theory}

Since we are working within the framework of a plane-parallel atmosphere,
the abundance distributions of elements that we are able to compute
necessarily comprise only vertical stratifications. However, it will be
possible to infer horizontal chemical inhomogeneities from plane-parallel 
results by varying the magnetic field angle with respect to the vertical. 
In this section we start by discussing vertical stratifications.

In the test-particle approximation, the 1D continuity equation for a 
given element may be written as
\begin{equation}
\partial _t N + \partial _z \left[ {N\left( {V_D  + V_M } \right)} \right] = 0~,
\label{eqCont}
\end{equation}
where $N$ is the local number density of the element in consideration and
$V_D$ its diffusion velocity; $V_M$ is a macroscopic velocity (a global 
flow of matter). $V_M$ is unknown, but in the present study we consider
only the case with no macroscopic motion ($V_M = 0$) \footnote{In the 
case of stationary mass loss, $V_M$ is the velocity of the stellar wind 
and can easily be determined for plane-parallel geometry,
noting that the flux of matter must be constant 
throughout the atmosphere.}. In our numerical computations, we use the 
full diffusion velocity and the diffusion flux, as given by Eqs.(4) and
(13) to (17) of Alecian \& Stift (\cite{als06}), but for the sake of clarity 
in this discussion, we adopt the following schematic expression for
the diffusion velocity: 
\begin{equation}
V_D  \approx \left\langle D \right\rangle \left[ { - \;\partial _z \ln 
\frac{N}{{N_H }}\; + \frac{{Am_p }}{{kT}}\left( {g^{rad}  - g} \right)} \right]
\label{VDs}
\end{equation}
where $\left\langle D \right\rangle$ is the average diffusion coefficient 
and $g^{rad}$ the radiative acceleration. All other symbols have their 
usual meanings.

The partial $z$-derivative in Eq.\,(\ref{VDs}) represents the 
\emph{ordinary} diffusion term, which is zero for homogeneous 
concentrations. When an abundance stratification appears in the 
atmosphere, this term introduces a Laplace operator in Eq.\,(\ref{eqCont})
which tends to smooth the concentration. In case of small
turbulent motions, this derivative must be multiplied by a 
coefficient to account for the mixing (Schatzman \cite{schatz69}). In 
practice, this term prevents the appearance of sharp-edged 
stratifications, but it does not strongly affect the global shape 
of the stratification profile and is not sensitive to an abundance offset.

Equation~(\ref{eqCont}) describes the evolution with time of the 
abundance of the element considered. Initial abundances are 
generally supposed to be solar; the starting time ($t = 0$) is 
supposed to correspond to the moment when mixing processes become 
negligibly small with respect to diffusion processes. The evolution 
of the abundances depends on boundary conditions (not discussed in
this paper). Because the diffusion velocity varies by several orders 
of magnitude from the bottom of the atmosphere to the highest layers, 
we are faced with a stiff problem. Atmospheres being optically thin, 
the calculation of radiative accelerations requires detailed non-local 
radiative transfer solutions for each wavelength and time step. This
leads to very expensive numerical computations which currently are
only carried out for stellar interiors which have the advantage of
being optically thick (see for instance Turcotte et al. \cite{tur98}, 
Seaton \cite{sea99}). As already mentioned, for stellar atmospheres, 
an accurate solution of Eq.\,(\ref{eqCont}) is still out of reach. However, 
one can propose approximate solutions for abundance stratifications.

\subsection{Equilibrium hypothesis}
\label{subsec:Equihyp}

Approximate solutions for abundance stratifications are often based 
on the following hypothesis. It is supposed that the time-dependent 
process described by  Eq.\,(\ref{eqCont}) builds up a stable 
stratification which fulfills $\partial _z 
\left[ {N\left( {V_D  + V_M } \right)} \right] = 0$.  This defines a 
class of stationary solutions (constant element flux throughout the 
atmosphere). Noting that we have set $V_M = 0$, a particular sub-class 
of solutions corresponds to the case $V_D = 0$, which we call \emph{equilibrium} 
solutions. In other words, an \emph{equilibrium} solution corresponds 
to an abundance stratification $N(z)$ such that $V_D = 0$ everywhere 
in the atmosphere. The advantage of looking for an \emph{equilibrium} 
solution is that it can be obtained by a simple iterative method and 
that it is much easier and cheaper to obtain than the correct solution
of Eq.\,(\ref{eqCont}). All studies of abundance stratifications in
stellar atmospheres published so far have been carried out under the
assumption of the equilibrium or (more rarely) of the stationary 
hypothesis. According to Alecian \& Grappin (\cite{alg84}), stationary
solutions exist in the optically thick case, but the optically thin 
case is still an open problem.

An important point is that stationary solutions are not solutions of the
continuity equation, since one imposes $\partial _t N = 0$. This means that
there is no particle conservation: for instance, one does not ensure that the
quantity of particles needed to obtain strong accumulations of an element at
some place in the atmosphere can actually be provided by adjacent layers. This
problem, when applied to the whole atmosphere, has been named in old works as
the ``reservoir" problem: are there enough particles coming from below the
atmosphere (the \emph{reservoir}!) to explain the observed over-abundances
of some elements in CP stars? Generally, the answer to this question has been
positive when the radiative acceleration of the element considered was found
significantly in excess of gravity at the bottom boundary layer (see
computations by Seaton, \cite{sea99}).

Two more fundamental questions arise when looking for equilibrium 
(or for stationary) solutions: does a solution always exist, and 
if it exists, is it unique? Formally, the answer to the first question 
is no. Neglecting the concentration gradient in Eq.\,(\ref{VDs}), the
equilibrium solution corresponds to $g^{rad} = g$. One knows that 
$g^{rad}$ generally has its maximum value at the limit of vanishing 
concentration and then decreases with increasing concentration. If
this maximum $g^{rad}$ is smaller than $g$, an equilibrium solution 
does not exist. This may be the case for elements with insufficient 
absorption capabilities at some depth in the stellar atmosphere. The 
answer to the second question is yes in the optically thick case, as 
long as the variation of $g^{rad}$ with respect to $N$ is 
monotonic\footnote{In self-consistent modelling, when several elements
stratify simultaneously, a variation in $N$ can affect the structure
of the star, the variation in $g_{rad}$ with respect to $N$ can be
non-monotonic, and nothing can be said about the uniqueness of the
solutions, even in the optically thick case.}. 
In the optically thin case, where the element concentration in some 
layer affects the radiation field in other layers, there is no reason 
to assert that this solution should be unique. We will come back to this 
point in Sect. \ref{sec:resdisc}.

\section{Numerics}
\label{sec:numerics}
From our previous papers (Alecian \& Stift \cite{als04}, \cite{als06}), it 
emerges that both accelerations and diffusion velocities vary strongly 
with depth in the stellar atmospheres. While radiative accelerations 
are not very sensitive to the direction of the magnetic field, 
horizontal magnetic fields can impede diffusive motions of ionised 
species quite effectively, in stark contrast to vertical fields. 
When looking for equilibrium stratifications, we therefore are 
potentially faced with large differences in chemical abundances 
between the different layers and with ensuing strong abundance 
gradients. The question arises of how to determine the equilibrium 
stratification by some reasonably simple iterative method, given
all the non-linearities inherent in radiative transfer and radiative
accelerations.

Fortunately, and rather unexpectedly, the problem proves to be numerically
user-friendly. Starting from a vertically homogeneous solar composition, we
first increase the abundance by some constant value throughout the atmosphere.
The resulting curve of diffusive flux or of effective acceleration\footnote{See
Eq.\,(15) of Alecian \& Stift (\cite{als06}) for the definition.} vs.
optical depth lies below the curve calculated with solar composition. The
respective abundance values required for either zero flux or for zero effective
acceleration at each depth point are now determined either by linear
interpolation or by (bounded) linear extrapolation. The new non-constant
abundance stratification obtained by this inter- or extrapolation constitutes
the basis for the next iteration step, which is carried out in exactly the same
way, possibly with smaller abundance offsets. Despite the slightly non-local
nature of the effects of abundance gradients, it is possible to adjust the
abundances point by point independently in depth without jeopardising
convergence.

Problems can arise with both convergence criteria used in this study.
Ideally, we would like to achieve zero diffusion flux throughout the 
atmosphere. Due to the high densities in the deeper layers, we can
reduce the original flux obtained with solar abundances by several
orders of magnitude, but the residual fluxes near the bottom of the
atmosphere can still exceed the fluxes near the surface by far
because of the very low densities there. Sometimes it is not possible 
to find an equilibrium stratification for the outermost layers and
abundances may drop by 5-10\,dex and more unless constrained to some
minimum value. This can lead to spurious humps in the stratification 
curve near the drop. 

The other criterion which we have extensively used in this study is 
zero effective acceleration. It is more efficient in the outer layers 
than the zero flux criterion, but it suffers from the drawback that it 
does not take the abundance gradients into account.
It turns out that it is not possible to give a general assessment of
the stability of the iteration procedure based on one or the other of 
the two convergence criteria. We also tried a combination of both 
and sometimes appear to achieve improved convergence. Still, 
instabilities are an annoying fact and a lot of calculations
diverge.

For a given element and a given atmosphere, our results 
have always converged towards the same final stratification, 
regardless of the initial abundances. Similarly, differences in 
equilibrium solutions are not significant between computations
carried out with different convergence criteria.

\section{Results and discussion}
\label{sec:resdisc}

Computations presented in this section were obtained with our new CaratStrat
code, which performs the iterative procedure described in Sect.
\ref{sec:numerics}. Diffusion velocities are calculated with CARAT, which is now
embedded in CaratStrat. Generally, about 10 iterations are needed to approach
equilibrium solutions of abundance stratifications for each element (each run of
the code adjusts the stratification of one element only). Sometimes, an
equilibrium solution cannot be reached in all parts of the atmosphere
($ -5.0 \le \log \tau \le 2.0$). However,
the results shown in this section are significant enough for discussion; places
where equilibrium is not reached are marked in the figures, together with an
indication of the trend directions. Since computations are quite expensive, we
restricted them (except for the $T_{\rm eff}=12\,000$\,K model) to elements
that have recently published observations and for which abundance stratifications
have been estimated from the observed spectra. We considered  models with
$T_{\rm eff}$ close to those of the observed stars, but in this study we did not
try to match {\em exactly} the published effective temperatures and gravities.
We have considered for each model the zero field case and 10\,kG magnetic fields
inclined under several angles. For the $T_{\rm eff}=10\,000$\,K model, we also present
computations for 1\,kG.

\subsection{Stellar atmospheres of different temperatures}
\label{subsec:atmos}

As in our previous papers (Alecian \& Stift \cite{als04}, \cite{als06}), we 
relied on Kurucz (\cite{kur93}) stellar atmospheric models with the tabulated 
continuous opacities required by CARAT taken directly from the 
output of ATLAS9 and corresponding to solar abundances. Similar
to the slight inconsistency concerning the accelerations from b-f transitions
discussed in Alecian \& Stift (\cite{als06}) -- cross sections are taken from
TopBase (Cunto \& Mendoza \cite{cum92}; Cunto et al. \cite{cum93}), 
continuous opacities for the flux calculations from ATLAS9 -- 
for some of the elements in our study we may encounter a discrepancy between
the continuous opacities derived with solar abundances and the continuous 
opacities corresponding to the actual stratifications. The results for 
species like Ti that do not contribute any conspicuous continua will hardly 
be affected, but Si could possibly constitute a different case. 
To clarify this issue, new versions of CARAT and CaratStrat have been
established which incorporate the entire ATLAS12 opacity package (for the latter
see Bischof 2005), allowing straightforward calculation of stratified
opacities. In this context one should keep in mind that, even without
stratification, there are large differences in the ``cool" Si\,{\sc i} and 
C\,{\sc i} opacities
between ATLAS9 and ATLAS12 in the UV. Therefore a comparison between the
respective Si equilibrium stratification obtained with the old and the new
versions of CARAT will reflect both the changes from ATLAS9 to ATLAS12 and the
influence of fixed vs. stratified continuous opacities. As it turns out,
differences in Si stratification between the old and the new versions do not
exceed 0.2 dex in those places where equilibrium is achieved. Given the
uncertainties in both theory and observations, these differences can be
considered marginal.

\subsection{Stratifications for $T_{\rm eff}=8\,500$\,K}
\label{subsec:strat8500}

\begin{figure}
\includegraphics[width=9.3cm]{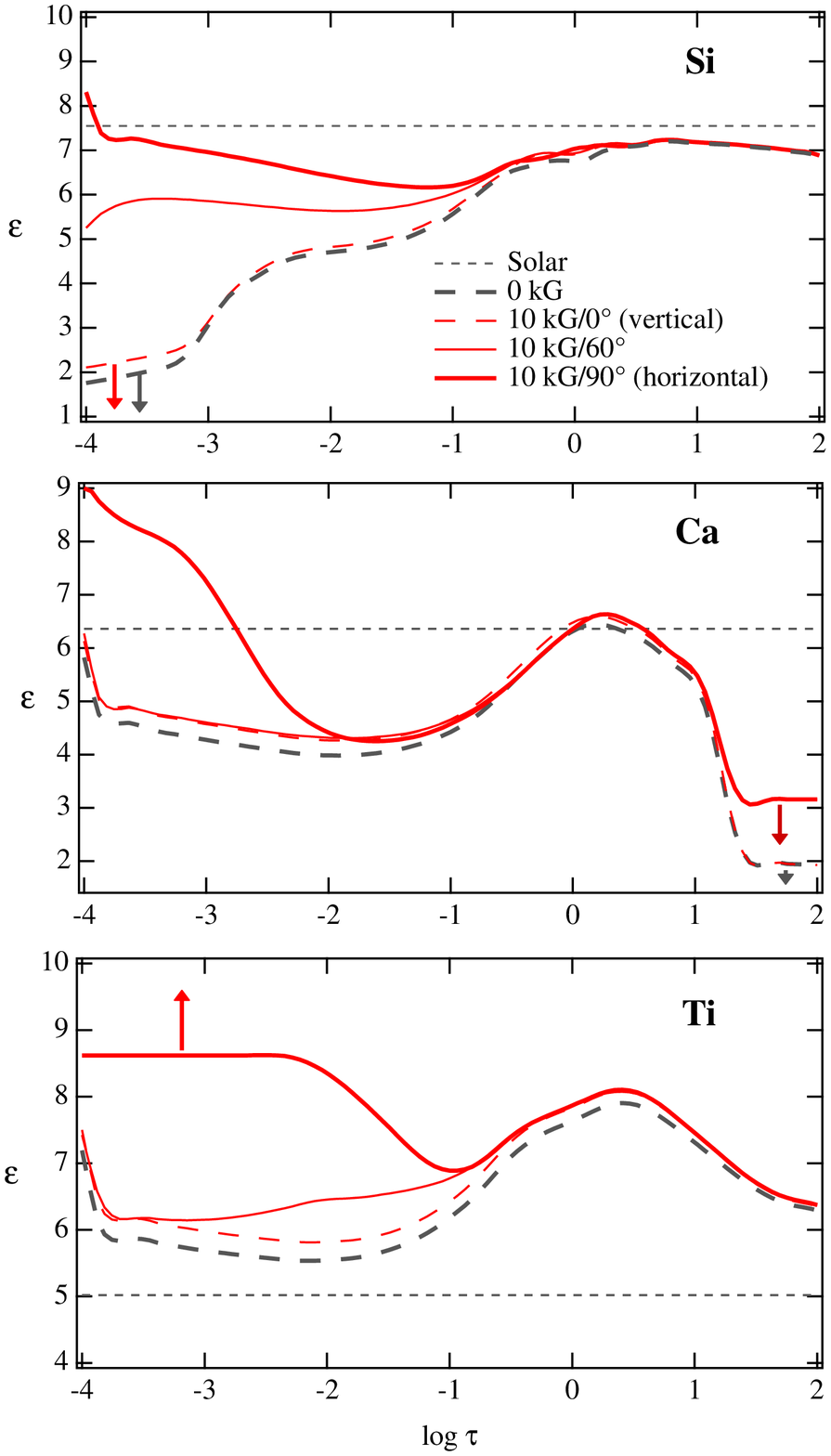}
\caption{
Abundance stratifications (equilibrium solutions) of Si, Ca, and Ti, for an
atmosphere with  $T_{\rm eff}=8\,500$\,K and $\log\,g=4.0$. The logarithm of 
the abundance ($\varepsilon$) with respect to hydrogen (in units of
$\log\,H=12$) is plotted against $\log \tau_{\rm 5000}$. The meaning of the 
different curves is given in the top panel (Si); the arrows indicate
that the equilibrium solution is not reached in some layers and that 
equilibrium abundances should be smaller (or larger) than those displayed.
}
\label{fig:Fig_strat8500lay}
\end{figure}

Equilibrium solutions for Si, Ca, and Ti are shown in
Fig.\,\ref{fig:Fig_strat8500lay}. We have plotted $\varepsilon$, which is the
abundance (logarithmic particle number density) with respect to H where
$\varepsilon (\rm H) = 12$. The effective temperature of $8\,500$\,K is close to
the value of $8\,400$\,K adopted by Kochukhov et al. (\cite{koc04}) for 53\,Cam,
but the gravity of the model we used is higher ($4.0$ instead of $3.7$, see
Sect.\,\ref{subsec:atmos}). The heavy-long-dashed and long-dashed lines are
stratifications for zero field and for a 10\,kG vertical magnetic field,
respectively. Differences between these two curves are thus only due to Zeeman
amplifications of the radiative accelerations \footnote{According to the usual
approximation, the diffusion velocity in a plane-parallel atmosphere is
insensitive to vertical magnetic fields.}. The effect of Zeeman amplification is
weak for Si, because Zeeman splitting of its strong absorption lines in the UV
is comparatively small; this is in accord with the findings of Alecian \& Stift
(\cite{als04}).

\begin{figure}
\includegraphics[width=9cm]{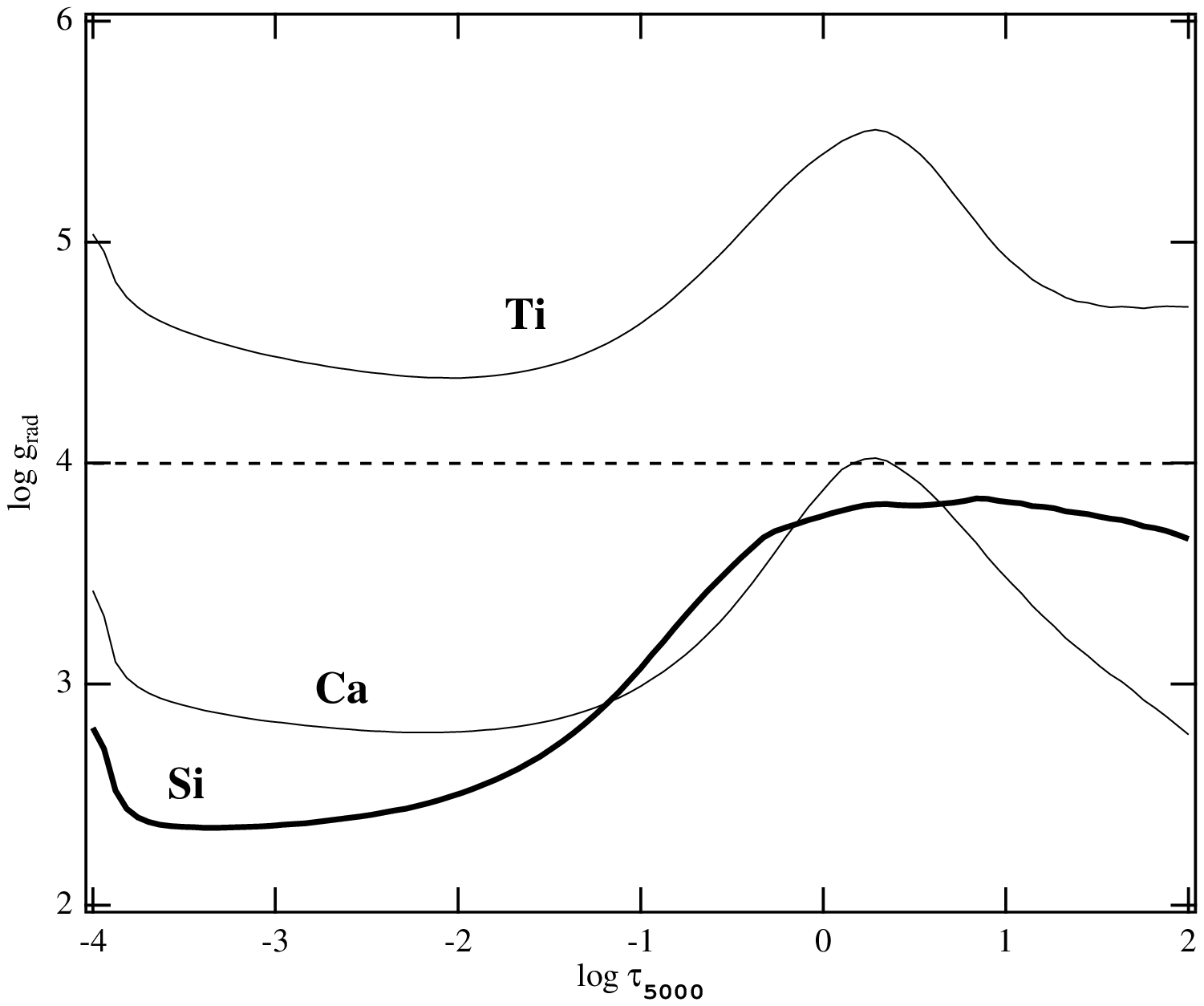}
\caption{
Radiative accelerations of Si, Ca, and Ti [$\log (cm\,s^{-2})$] for a zero 
magnetic field and solar homogeneous abundances in the atmospheric model 
of Fig.\,\ref{fig:Fig_strat8500lay}. The dashed line indicates the gravity.
}
\label{fig:grad8500lay}
\end{figure}

The role of the magnetic field remains minor for layers deeper than $\log \tau =
-1.0$, because the collision rates increase with particle density (see the
discussion on diffusion coefficients in Alecian \& Stift \cite{als06}, and
their Fig.\,4c). Equilibrium stratifications in these layers are -- as expected
-- closely related to radiative accelerations for solar homogeneous abundances, as
shown in Fig.\,\ref{fig:grad8500lay}, since these layers are optically thick: Si
is not supported by the radiation field for solar abundance, Ca is barely
supported around  $\log \tau = 0.0$, and Ti is strongly pushed upwards unless it
becomes strongly overabundant. No equilibrium solution is obtained for Ca in
layers deeper than $\log \tau = 1.5$; Ca is in a noble gas configuration in
these layers and not enough photons are absorbed to support Ca. Consequently,
the Ca abundance must decrease at the bottom of the atmosphere towards a value
that is probably lower than the one shown in the figure, indicated by the downward
arrows. The final Ca deficiency of the bottom layers of the atmosphere will be
determined by the diffusion flux in the envelope and the abundance contrast
allowed by the concentration gradient term.

For layers higher than $\log \tau = -1.0$, the magnetic field is very efficient,
especially when it is horizontal. This is generally due to the fact that
element diffusion in the neutral state is favoured by the presence of a
horizontal component of the magnetic field. Generally, the neutral state
undergoes strong radiative acceleration (absorption lines are not saturated) and
has a large diffusion coefficient. Silicon is still hardly supported in a
horizontal field, but much better than in a vertical field. Calcium and titanium
abundances are clearly enhanced above  $\log \tau = -2.0$ for a horizontal
field.

According to the surface abundance patterns reconstructed by Kochukhov et al.
(\cite{koc04}) for 53\,Cam (their Fig.\,10), Si is more abundant in places where
the magnetic field is nearly horizontal. This is consistent with the trend shown
in Fig.\,\ref{fig:Fig_strat8500lay}, but our predicted absolute abundances of Si
are at variance with the maps, being always smaller than the solar value. This
could be due to the higher gravity in our model than found for 53\,Cam.
Calcium is strongly stratified in our computations, with an enhancement only in
layers above $\log \tau = -3.0$ in the presence of a strong horizontal magnetic
field, but the correlation with the magnetic map of 53\,Cam is less clear than
in the case of Si. The Ti abundance map appears to be more or less
anti-correlated with the Ca distribution, except near the visible magnetic pole.
This anti-correlation could possibly be related to the \emph{cloud} of Ti we
find for a horizontal field about $\log \tau = -2.0$ and the \emph{hole} of Ca
we find in the same layers. At the magnetic pole, both Ca and Ti are enhanced in
53\,Cam. In Fig.\,\ref{fig:Fig_strat8500lay}, Ca and Ti equilibrium
stratifications near the magnetic poles can be deduced from the dashed lines.
One notices that both curves have very similar profiles, apart from their
distance to the respective solar values. Does this similarity explain the
enhancements of Ca and Ti at the visible pole of 53\,Cam? One cannot make this
assertion from our results. It could also be the signature of mass loss at the
magnetic pole in combination with the diffusion velocity (see the discussion by
Babel \& Michaud, \cite{bam91}), and in that case, an equilibrium solution
cannot account for this kind of situation.

\subsection{Stratifications for $T_{\rm eff}=10\,000$\,K}
\label{subsec:strat10000}

\begin{figure*}
\includegraphics[width=18.5cm]{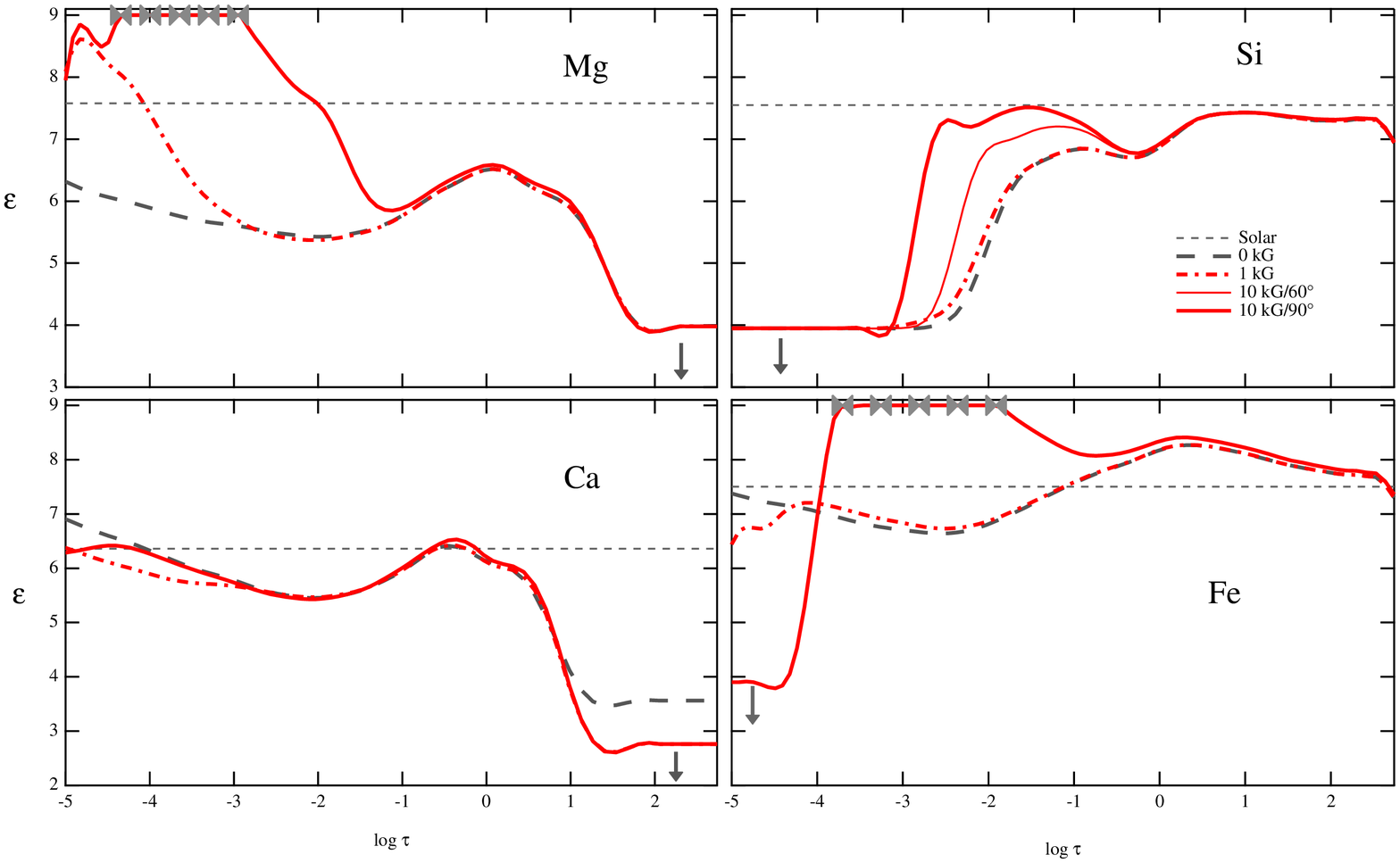}
\caption{
Abundance stratifications (equilibrium solutions) of Mg, Si, Ca, and Fe, for an
atmosphere with  $T_{\rm eff}=10\,000$\,K. The dashed-point lines correspond to
the case of a 1\,kG horizontal magnetic field, for the other curves the legends
are the same as in Fig.\,\ref{fig:Fig_strat8500lay} (see text for details).
}
\label{fig:Fig_strat10000lay}
\end{figure*}

Equilibrium solutions for Mg, Si, Ca, and Fe are shown in
Fig.\,\ref{fig:Fig_strat10000lay}. The effective temperature of $10\,000$\,K is
higher than the one adopted by Kochukhov et al. (\cite{koc06}) for HD\,133792
($9\,400\pm 200$\,K), and the gravity of the model we used is also larger ($4.0$
instead of $3.7$). However, these values are close enough to allow a comparison
of our results with the stratifications observationally deduced for this star.

\begin{figure}
\includegraphics[width=9cm]{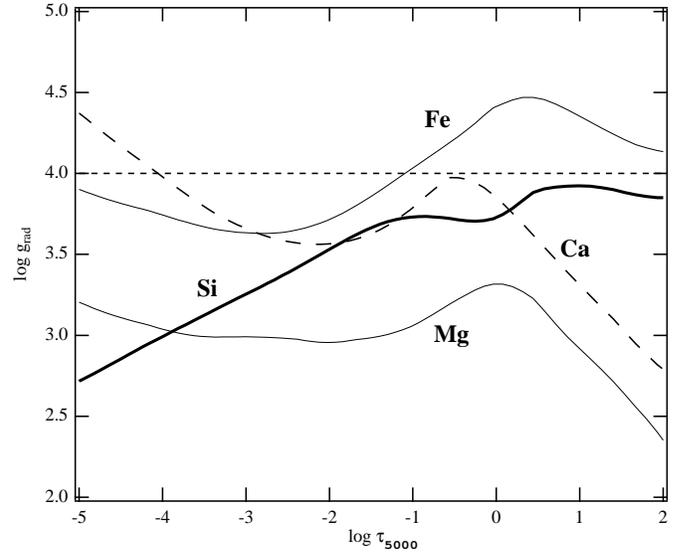}
\caption{
Radiative accelerations of Mg, Si, Ca, and Fe [$\log (cm.s^{-2})$] for zero 
magnetic field and solar homogeneous abundances in the atmospheric model of
Fig.\,\ref{fig:Fig_strat10000lay}. The dashed line indicates the gravity.
}
\label{fig:grad10000lay}
\end{figure}

For Mg, Ca, and Fe, the magnetic cases with angles 0 and 60 degrees are not shown
because they cannot really be distinguished from the non-magnetic case. One
notices that for Si and Ca, equilibrium solutions are clearly different from
those found for $T_{\rm eff}=8\,500$\,K; the same is true for radiative
accelerations (compare Fig.\,\ref{fig:grad10000lay} to
Fig.\,\ref{fig:grad8500lay}). The plateau for Mg and Fe (in a 10\,kG field)
around $\log \tau = -3.0$ is due to the fact that we limit the abundances to
$\varepsilon \le 9.0$\,. Larger abundances could be unphysical since our
computations are done in the test-particle approximation and for an atmospheric
model with solar abundance. For this particular model, we also considered
the case of a moderate horizontal field strength of 1\,kG (close to the average
field observed in HD\,133792). One can see that its effect is non-negligible for
Mg. This is due to the strong contribution (enhanced by the horizontal magnetic
field) of Mg\,{\sc i} to the diffusion velocity.

We can compare these stratifications to those derived from high resolution spectra of
HD\,133792 by Kochukhov et al. (\cite{koc06}) and shown in their Fig.\,5. For Mg, they
find a strong accumulation (about 2\,dex) above $\log \tau = -3.0$. Our
equilibrium solution shows the same kind of enhancement for 1\,kG, and a
deficiency of Mg around $\log \tau = -2.0$. The contrast between overabundant
layers and underabundant layers is also about $10^2$, but our equilibrium
solution is more complex and predicts a Mg \emph{cloud} that is more pronounced
and goes deeper for a 10\,kG horizontal field.

For Si, our equilibrium solution gives a decrease in the Si abundance above
$\log \tau = 0.5$ and a very strong depletion (3\,dex) in layers higher than
$\log \tau = -2.0$. Again, the contrast is comparable to what is deduced from
observed spectra, but the drop in Si abundance occurs higher up in the
atmosphere in our solution, the height of the drop increasing with the strength
of the magnetic field.

For Ca, our equilibrium solution is quite different from the stratification
proposed by Kochukhov et al. (\cite{koc06}), since the equilibrium solution
shows a moderate decrease in the Ca abundance around  $\log \tau = -2.0$
and not a step-function-like decrease like the one derived from the spectral
analysis. On the other hand, Ca is strongly underabundant below $\log \tau =
0.5$ in the equilibrium solution (as in the case of the $T_{\rm eff}=8\,500$\,K
model), because in these layers Ca is mainly in noble gas configuration
(Ca\,III) for which the radiative acceleration is weak. We doubt that the
spectral analysis used by Kochukhov et al. (\cite{koc06}) can reliably diagnose
these optically thick deep layers.

For Fe, the equilibrium solution for a horizontal field of $< 1$\,kG reveals a
contrast of about 1.5\,dex between overabundant layers (around $\log \tau =
0.0$) and deficient layers (above $\log \tau = -2.0$). This is of the same order
of magnitude as what has been observationally derived for HD\,133792. The
transition between overabundant and deficient layers occurs more or less at the
same depth in the empirical profile and in our equilibrium solution,
but the stratification profile is not monotonic in the latter and does not
exhibit, as the empirical profile does, a sharp drop above $\log \tau = -2.5$
from a strong enhancement below.

\subsection{Stratifications for $T_{\rm eff}=12\,000$\,K}
\label{subsec:strat12000}

\begin{figure}
\includegraphics[width=9.3cm]{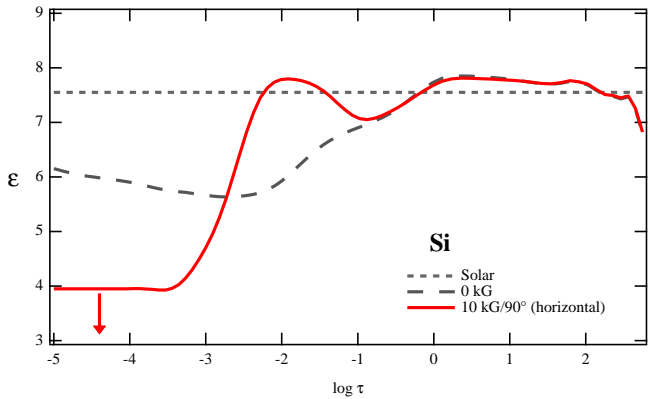}
\caption{
Abundance stratifications (equilibrium solutions) of Si, for an atmosphere with 
$T_{\rm eff}=12\,000$\,K (same legend as Fig.\,\ref{fig:Fig_strat10000lay}). In
the magnetic case, some layers around $\log \tau = -2.0$ could exhibit an 
overabundance of Si. However, according to the accelerations shown in
Fig.\,\ref{fig:grad12000lay} (long-dashed curve) Si is not supported above $\log
\tau = 0.0$ for solar abundance. The local overabundance is due to the 
deficiency of Si below these layers: the medium being optically thin, more
photons come from deficient layers and support silicon in the layers above.
}
\label{fig:Fig_strat12000lay}
\end{figure}

\begin{figure}
\includegraphics[width=9cm]{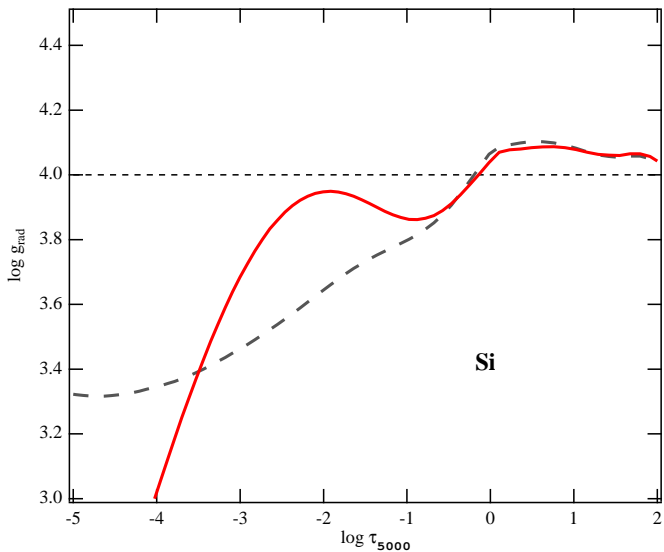}
\caption{
Radiative accelerations of Si [$\log (cm.s^{-2})$] for homogeneous solar 
abundance and the atmospheric model of Fig.\,\ref{fig:Fig_strat12000lay}. 
The dashed line indicates the gravity. The long-dashed curve is the acceleration 
for zero magnetic field, and the solid line is the total acceleration in the
magnetic case (as defined in Alecian \& Stift, \cite{als06}, Eq.\,13).
}
\label{fig:grad12000lay}
\end{figure}

We also had a look at the behaviour of silicon in a hotter model, although we
have not found corresponding recent observations with stratification analysis.
This element is known to be enhanced in magnetic Ap stars but not in HgMn stars.
According to Fig.\,\ref{fig:Fig_strat12000lay}, the equilibrium solution we have
found  is consistent with such observations. Silicon clearly appears to be deficient
above $\log \tau = -1.0$ in the non-magnetic case. But, with a 10\,kG horizontal
magnetic field, a \emph{cloud} of Si forms near  $\log \tau = -2.0$, which should
lead to an apparent overabundance. It is to be noted that, according to the
accelerations shown in Fig.\,\ref{fig:grad12000lay}, a solar abundance of Si is
not supported by the radiation field around $\log \tau = -2.0$, even for a
10\,kG field! At first sight, this would appear to be at variance with the
\emph{cloud} found in Fig.\,\ref{fig:Fig_strat12000lay}. In fact, this is not so:
due to the deficiency of Si around $\log \tau = -1.0$ (just below the
\emph{cloud}), more photons are available than in the solar case to support the
Si \emph{cloud}. This is a typical behaviour in the optically thin case, which
cannot be encountered in stellar interiors.

\section{Conclusions}
\label{seq:conclusion}

In this work we present the first detailed numerical calculations of element
stratifications due to atomic diffusion in magnetic atmospheres. This study
addresses the abundance anomalies observed in Ap stars, either magnetic or
non-magnetic (HgMn stars). Radiative accelerations are based on full
opacity sampling of Zeeman split spectral lines, and diffusion velocities 
are obtained by the methods described in Alecian \& Stift (\cite{als06}). The
stratifications derived in the present study correspond to equilibrium
solutions: one looks for abundance stratifications such that diffusion
velocities are close to zero everywhere in the atmosphere.

Because of the high computational cost, we restricted our study to a few
metals and stellar atmospheres. We want to explore those cases where our results
can be compared with observed abundance maps and stratifications. We did not
try to use atmospheric models corresponding perfectly to those derived from
the observations of 53\,Cam and of HD\,133792, but instead used standard solar abundance
ATLAS9 models that are close enough to allow a meaningful discussion. It emerges
from this comparison that, in several cases, equilibrium solutions are consistent
with stratifications reconstructed from observed Stokes spectra by means of
magnetic Doppler imaging. However, significant differences also exist. Some of
these differences may be partly due to possible shortcomings in the inversion
of the observational material, but one has to keep in mind that our modelling
approach, with its search of equilibrium stratifications, is more likely to be
responsible. Indeed, equilibrium solutions can be very different from the stratifications
encountered in real stars, which result from non-linear, time-dependent
processes, with a competition between several physical processes (atomic
diffusion, inhomogeneous mass-loss, complex magnetic geometries, NLTE effects,
turbulence, etc.). The present study constitutes one more step toward the
detailed modelling of magnetic atmospheres, but for the first time in this
quest, the results we obtain can be compared with observations. Time-dependent
atomic diffusion is the next step, which we shall address in the near future.

\begin{acknowledgements} GA acknowledges the financial support of the Programme
National de Physique Stellaire (PNPS) of CNRS/INSU, France. MJS acknowledges
support by the {\sf\em Austrian Science Fund (FWF)}, project P16003-N05
``Radiation driven diffusion in magnetic stellar atmospheres'' and through a
Visiting Professorship at the Observatoire de Paris-Meudon and Universit{\'e}
Paris 7 (LUTH). Thanks go to AdaCore for generously providing us with the
GNAT\,Pro Ada95 compiler and toolsuite. A large part of the calculations were
carried out on the Sgi Origin\,3800 and IBM Power4 of the CINES in Montpellier.
\end{acknowledgements}

{}

\end{document}